\documentclass[prc,aps,twocolumn,showpacs,nofootinbib,superscriptaddress]{revtex4-1}
\usepackage{amsmath}
\usepackage{amssymb}
\usepackage{amsfonts}
\usepackage{graphicx}
\usepackage{color}
\usepackage{tabularx}

\renewcommand{\vec}[1]{{\mathbf #1}}
\newcommand{\nuc}[2]{$^{#1}${#2}}

\newcommand{\br}{\boldsymbol{\mathbf r}}

\newcommand{\bJ}{\boldsymbol{\mathbf J}}
\newcommand{\bW}{\boldsymbol{\mathbf W}}
\newcommand{\vnabla}{\boldsymbol{\mathbf\nabla}}

\newcommand{\dbr}{\mathrm d^3r}

\newcommand{\avol}{a_{\rm vol}}
\newcommand{\asurf}{a_{\rm surf}}
\newcommand{\acurv}{a_{\rm curv}}
\newcommand{\asym}{a_{\rm sym}}

\newcommand{\etal}{\emph{et al.}}
\newcommand{\nn}{\nonumber}

\begin{document}

\title{Constraining the surface properties of effective Skyrme interactions}

\author{R. Jodon}
\affiliation{IPNL, 
             Universit{\'e} de Lyon, Universit{\'e} Lyon 1, CNRS/IN2P3, 
             F--69622 Villeurbanne, France}

\author{M. Bender}
\email{bender@ipnl.in2p3.fr}
\affiliation{IPNL, 
             Universit{\'e} de Lyon, Universit{\'e} Lyon 1, CNRS/IN2P3, 
             F--69622 Villeurbanne, France}
\affiliation{CENBG, 
             Universit{\'e} Bordeaux 1, CNRS/IN2P3, 
             F--33175 Gradignan, France}

\author{K. Bennaceur}
\email{bennaceur@ipnl.in2p3.fr}
\affiliation{IPNL, 
             Universit{\'e} de Lyon, Universit{\'e} Lyon 1, CNRS/IN2P3, 
             F--69622 Villeurbanne, France}
\affiliation{Department of Physics, PO Box 35 (YFL),
             FI--40014 University of Jyv\"askyl\"a, Finland}
\affiliation{Helsinki Institute of Physics, P.O. Box 64,
             FI--00014 University of Helsinki, Finland}

\author{J. Meyer}
\email{jmeyer@ipnl.in2p3.fr}
\affiliation{IPNL, 
             Universit{\'e} de Lyon, Universit{\'e} Lyon 1, CNRS/IN2P3, 
             F--69622 Villeurbanne, France}
%
%

\begin{abstract}
\begin{description}

\item[Background]
Deformation energy surfaces map how the total binding energy of a
nuclear system depends on the geometrical properties of intrinsic
configurations, thereby providing a powerful tool to interpret 
nuclear spectroscopy and large-amplitude collective motion phenomena 
such as fission. The global behavior of the deformation energy is known to 
be directly connected to the surface properties of the effective 
interaction used for its calculation. 

\item[Purpose]
The precise control of surface properties during the parameter adjustment 
of an effective interaction is key to obtain a reliable and predictive 
description of nuclear properties. The most relevant indicator is the 
surface energy coefficient $\asurf$. There are several possibilities 
for its definition and estimation, which are not fully equivalent and 
require a computational effort that can differ by orders of magnitude. 
The purpose of this study is threefold: 
first, to identify a scheme for the determination of $\asurf$ that offers 
the best compromise between robustness, precision, and numerical efficiency;
second, to analyze the correlation between values for $\asurf$ and the
characteristic energies of the fission barrier of \nuc{240}{Pu}; and 
third, to lay out an efficient and robust procedure how the deformation 
properties of the Skyrme energy density functional (EDF) can be constrained 
during the parameter fit.

\item[Methods]
There are several frequently used possibilities to define and calculate 
the surface energy coefficient $\asurf$ of effective interactions built 
for the purpose of self-consistent mean-field calculations. The most 
direct access is provided by the model system of semi-infinite nuclear 
matter, but $\asurf$ can also be extracted from the systematics of binding 
energies of finite nuclei. Calculations can be carried out either 
self-consistently (HF), which incorporates quantal shell effects, or in one of 
the semi-classical Extended Thomas-Fermi (ETF) or Modified Thomas-Fermi (MTF)
approximations. The latter is of particular interest as it provides 
$\asurf$ as a numerical integral without the need to solve self-consistent
equations. Results for semi-infinite nuclear matter obtained with the HF,
ETF, and MTF methods will be compared with one another and with $\asurf$ as
deduced from ETF calculations of very heavy fictitious nuclei.

\item[Results]
The surface energy coefficient of 76~parameterizations of the Skyrme EDF 
have been calculated. Values obtained with the HF, ETF and MTF methods are 
not identical, but differ by fairly constant systematic offsets.
By contrast, extracting $\asurf$ from the binding energy of semi-infinite 
matter or of very large nuclei within the same method gives the same result 
within the numerical uncertainties.

\item[Conclusions]
Despite having some drawbacks compared to the other methods studied here, 
the MTF approach provides sufficiently precise values for $\asurf$ such that 
it can be used as a very robust constraint on surface properties
during a parameter fit at negligible additional cost. While the excitation 
energy of superdeformed states and the height of fission barriers is
obviously strongly correlated to $\asurf$, the presence of shell effects 
prevents a one-to-one correspondence between them. As in addition the value 
of $\asurf$ providing realistic fission barriers depends on the choices made 
for corrections for spurious motion, its ``best value'' (within a given scheme 
to calculate it) depends on the fit protocol. Through the construction of 
a series of eight parameterizations SLy5s1-SLy5s8 of the standard Skyrme 
EDF with systematically varied $\asurf$ value it is shown how to arrive
at a fit with realistic deformation properties.

\end{description}
\end{abstract}

%
%


\date{3 June 2016}

\maketitle

%
%
\section{Introduction}
\label{sect:intro}

Energy density functional (EDF) methods are versatile tools for the 
study of nuclear structure and dynamics \cite{RMP}. Once a parameterization 
of the EDF has been constructed by selecting 
terms that incorporate the relevant degrees of freedom and by carefully 
fixing each term's coefficient, it can be applied to nuclei throughout 
the chart of nuclei at the level of static or time-dependent mean-field 
calculations, random phase approximation (RPA) and its extensions, 
or for the description of large-amplitude motion in the context of the 
generator coordinate method.

One popular example is the widely-used Skyrme 
EDF~\cite{Sky56a,Sky58a,VB72,RMP}.
Its further improvement is presently pushed into two
major directions. One concerns the protocol for the adjustment of its
parameters, where the number and diversity of data and pseudo-data considered 
during the fit is increased and various post-fit analyses
added that allow for the quantification of confidence intervals of model 
parameters and the estimate of statistical error 
bars~\cite{Fri86a,Kortel10,Kortel12,gao13a,Kortel14,Doba14}.
The other is the set-up of more general forms of the Skyrme EDF 
containing higher-order terms with additional 
parameters~\cite{papI,Car08a,Raimondi11,Dav13a,Bec15a,Sadoudi_Kaz12,Sadoudi13}.
As both developments substantially increase the numerical cost of the 
parameters' adjustment, it is of advantage to have efficient methods to 
calculate the data and pseudo-data used during the fit.

In the past, many authors have discussed the crucial role that an accurate
adjustment of the surface energy coefficient of an EDF's parameterization 
plays for the deformation properties of 
nuclei~\cite{Bar82a,Berger91,Ben00a,Goriely07,Nik11,Kortel12,Kortel14}. 
Generally speaking, when increasing $\asurf$,
deformation energy surfaces become stiffer, 
fission barriers higher and the excitation energy of fission isomers and
other superdeformed states more elevated. 
This, in turn can then be used to fine-tune the EDF.
First pioneering calculations of fission properties indicated that the two 
early Skyrme parameterizations SIII~\cite{Bei75a} and SkM~\cite{skm} give 
a fission barrier height for \nuc{240}{Pu} that is too high or too low by 
roughly a factor of two, respectively. 
This disagreement clearly exhibited within semi-classical
calculations~\cite{Brack85} has triggered the
adjustment of the SkM$^*$ parameterization, for which the parameters of the
momentum-dependent part of SkM have been modified and fine-tuned in order to 
change the height of the semi-classical fission barrier without changing
other properties of infinite nuclear matter~\cite{Bar82a}. 
Even after more than 30 years, SkM$^*$ is a still often used reference for 
the description of fission 
phenomena~\cite{Sta13,staszczak13a,Schunck14a,sadhukhan13a,sadhukhan14a,mcdonnell14a}.

The aim of the present study is threefold.
First, we will set up and benchmark an efficient and robust method to 
calculate the surface energy coefficient $\asurf$ as defined in 
Ref.~\cite{cote78} for modern Skyrme EDFs. Second, we will analyze
the correlation between the value for $\asurf$ and characteristic 
energies of the deformation energy landscape of \nuc{240}{Pu}. 
Third, we will outline how a constraint on the value of $\asurf$ 
can be incorporated into the parameter adjustment.

This article is organized as follows. 
In Section~\ref{sect:semi-inm} we present the three methods
based on a one-dimensional model for semi-infinite nuclear matter
that will be used to evaluate the surface energy coefficient of the
Skyrme EDF: 
the self-consistent Hartree-Fock method (HF), 
the extended Thomas-Fermi method (ETF),
and the Modified Thomas-Fermi method (MTF).
Section~\ref{sect:nuclei} recalls the ``leptodermous protocol''
of Reinhard \etal~\cite{Reinhard06} to extract the surface energy
from an analysis of very large fictitious nuclei. We will use this 
approach with the ETF method.
A systematic comparison of the results provided by these four possibilities
is presented in Section~\ref{sect:results}. 
In Sect.~\ref{sect:fis}, we discuss to what extent the 
such extracted surface energy coefficient is
linked to characteristic energies of the fission barrier of \nuc{240}{Pu}.
Finally, Section~\ref{sect:sly5sx} presents the adjustment of a series 
of standard Skyrme parameterizations with systematically varied $\asurf$,
called SLy5s1--SLy5s8.
A summary of our discussion and perspectives will be given in 
Section~\ref{sect:conclusion}.

%
%
\section{Surface energy in semi-infinite nuclear matter}
\label{sect:semi-inm}

%
%
\subsection{The Skyrme EDF}
\label{sect:EDF}

The starting point for the derivation of a Skyrme EDF is usually an effective 
two-body interaction with parameters that may depend on the density of the 
system. It is important to recall that this effective interaction is only 
used as a generator for the general form of the EDF and allows for writing 
the coupling constants that weight these scalars as functions of a limited 
number of parameters.
It is then common to disregard certain terms in the functional or to relax 
some of the interdependences between the coupling constants, which breaks 
the one-to-one-correspondence between the EDF and the underlying effective 
interaction used to construct it \cite{RMP,papI,Sadoudi13,Ryssens14}. 

The present standard form of the Skyrme EDF is motivated by the use
of a density-dependent two-body interaction \cite{RMP}, leading to a
bilinear EDF with density-dependent coupling constants. Most of the
parameterizations discussed below will be of that type. However, we 
will also compare with results obtained for EDFs derived from interactions 
with two-, three- and sometimes even four-body terms but density-independent 
coupling constants \cite{Sadoudithesis,Sadoudi_Kaz12,Sadoudi13}. Such
functionals will be referred to as trilinear or quartic EDFs, respectively.

In general, the total energy can be written as the sum of five 
terms~\cite{RMP}:
the kinetic energy $E_\mathrm{kin}$,
a potential energy functional $E_\mathrm{Sky}$ that models the strong
interaction in the particle-hole channel,
a pairing energy functional $E_\mathrm{pairing}$,
a Coulomb energy functional $E_\mathrm{Coulomb}$
and a correction term $E_\mathrm{corr}$ that approximately removes
the excitation energy due to spurious motions caused by broken symmetries
\begin{equation}
\label{eq:efu:complete}
E = E_{\text{kin}}
  + E_{\text{Sky}}
  + E_{\text{pairing}}
  + E_{\text{Coulomb}}
  + E_{\text{corr}} \, .
\end{equation}
The model systems that we will use to extract the surface properties 
are semi-infinite nuclear matter and giant spherical nuclei,
for which only the first two parts of the energy are taken into account.
As a consequence of the Skyrme interaction being a contact force, the
corresponding EDF then can be written in the form of an integral
over a local energy density
\begin{eqnarray}
\label{eq:edf:sk}
E 
& = & \int \! \dbr \, \mathcal{E} (\br) \, ,
      \\
\mathcal{E} (\br) 
& = &   \mathcal{E}_\mathrm{kin} (\br)
      + \mathcal{E}_\mathrm{Sky} (\br) \, ,
\end{eqnarray}
where the Skyrme part $\mathcal{E}_\mathrm{Sky} (\br)$
can be further decomposed into central ($\mathcal{E}^\mathrm{C}_t$),
spin-orbit ($\mathcal{E}^\mathrm{LS}_t$) and tensor 
($\mathcal{E}^\mathrm{T}_t$) terms 
\begin{eqnarray}
\mathcal{E}_\mathrm{Sky} (\br) 
& = & \sum_{t=0,1} \big[ \mathcal{E}^\mathrm{C}_t  (\br)
                          + \mathcal{E}^\mathrm{LS}_t (\br)
                          + \mathcal{E}^\mathrm{T}_t  (\br)
                   \big]                                   \, ,
\label{eq:edf:skb}
\end{eqnarray}
that are either composed entirely of isoscalar densities $(t=0)$ or
that contain bilinear combinations of isovector densities $(t=1)$. The
Skyrme energy functional as such is constructed to be an isoscalar.

The physics contained in the Skyrme functional has been discussed in
great detail in the literature~\cite{RMP,Per04a,papI,Sadoudi13,Ryssens14},
and here we will use standard notations for local densities and 
coefficients of the functional. Its complete form discussed in these
papers contains many terms, only a small subset of which are present for the 
systems we consider here. In particular, all time-odd densities are
zero because of the time-reversal invariance we impose, and many 
components of the time-even vector and tensor densities are zero 
because of spatial symmetries.

%
\subsection{Model of semi-infinite nuclear matter}

The surface energy is often extracted from an idealized one-dimensional
model of semi-infinite nuclear matter originally developed by 
Swiatecki~\cite{Swiatecki51} and revisited in~\cite{cote78}.
One considers a medium where the local densities are constant along the
$x$ and $y$ directions, but vary along the $z$ direction. The corresponding
profile of the local proton and neutron matter densities
will be noted as $\rho_q(z)$, $q= \text{n}$, p. Deep inside the matter for 
$z \to -\infty$, one expects that 
$\rho_0(z) \rightarrow \rho_\mathrm{sat}$, i.e.\ 
the equilibrium density of infinite nuclear matter, and that
${\cal E}(z) \to a_v$, i.e.\ the volume energy per particle at saturation. 
Far outside the matter, i.e.\ for $z \to +\infty$, one has $\rho_0(z) \to 0$.
For sake of compact notation, we will drop the $z$-dependence of the
densities from hereon whenever possible.

Within this one-dimensional model, the only non-vanishing components 
of the cartesian spin-current tensor density $J_{0,\mu \nu}$ are 
$J_{0,xy} = - J_{0,yx}$. As a consequence, only the $J_{0,z}$ component of 
the vector part ${\bJ}_0$ of the full spin-current tensor as defined 
in \cite{Per04a,papI} is non-zero.

We will here focus on the discussion of the surface energy coefficient 
$\asurf$ that is related to the surface energy of \textit{symmetric} 
semi-infinite nuclear matter. In this system, proton and neuton densities 
are equal, i.e.\ $\rho_n(z) = \rho_p(z) = \tfrac{1}{2} \rho_0(z)$ and 
similar for the other densities. As a consequence, only those terms in 
the EDF that are entirely composed of isoscalar densities have to be 
considered. This enormously simplifies the energy density 
\eqref{eq:edf:sk} that can be reduced to\footnote{Note that no unique
definition of the coupling constant $C^J_t$ of the tensor terms  
can be found in the literature. We use here the convention of 
Ref.~\cite{papI}. Others might differ by a factor two.
}
\begin{eqnarray}
\label{eq:esk_sym}
\mathcal{E}
& = & \frac{\hbar^2}{2m^*_0[\rho_0]} \, \tau_0
      \nn \\
&   & + C_0^\rho[\rho_0] \rho_0^2
      - C_0^{\Delta \rho} \big( \vnabla \rho_0 \big )^2
      + \tfrac{1}{2}\, C_0^J \bJ_0^2
      - C_0^{\nabla J} \bJ_0 \cdot \vnabla \rho_0 
      \nn \\
&   & + B^{\rho}_0 \rho_0^3
      + B^{\nabla \rho}_0 \rho_0 \big( \vnabla \rho_0 \big)^2
      + \tfrac{1}{2}\, B^J_0 \bJ_0^2 \rho
      + D^{\rho}_0 \rho_0^4 \, ,
\end{eqnarray}
where the density-dependent isoscalar effective mass is given by the ratio
\begin{equation}
\label{eq:f_0}
\frac{m}{m^*_0[\rho_0]}
= 1 \, + \, \frac{2m}{\hbar^2} \,
  \big( C_0^\tau \rho_0 + B^{\tau}_0 \rho_0^2 \big) \, .
\end{equation}
The expression for the Skyrme EDF provided by Eq.~\eqref{eq:esk_sym} covers 
a vast number of different parameterizations that have been lately used in 
the literature.
We use a notation where the coupling constants for the bilinear, trilinear
and quartic terms are denoted with $C_0$, $B_0$ and $D_0$, respectively. 

In the most-widely used standard form of the Skyrme EDF only the bilinear
terms are considered, i.e.\ all $B_0 = D_0 = 0$, and the coupling constant
$C_0^\rho$ is made explicitly density dependent by multiplying it with
$[ 1 + c \, \rho_0^\alpha(\vec{r})]$, where the parameter $c$ controls
the relative weight of the density-dependent part of the coupling constant.

The $\bJ_0^2$ terms bilinear in the spin-current density will 
be called ``tensor terms'' in what follows. For a majority of the 
widely-used parameterizations the coupling constant $C_0^J$ of the tensor 
terms is set to zero, such that $\mathcal{E}^{\mathrm{T}}_t = 0$ in 
Eq.~\eqref{eq:edf:skb}. Many past semi-classical and HF calculations of 
semi-infinite nuclear matter, however, have neglected these terms also
for those parameterizations for which they are to be taken into account.

The possibility of replacing the density dependence of $C^\rho_0$ by 
trilinear ($B_0 \neq 0$) and quartic ($D_0 \neq 0$) terms, and where 
all coupling constants are derived from an underlying Skyrme 2+3+4-body
Hamiltonian, has been considered recently with the goal of constructing
well-defined EDFs for use in beyond-mean-field methods 
\cite{Sadoudi_Kaz12,Sadoudi13}. This extended form of the Skyrme EDF 
will also be considered in the present work.

The assumptions made when setting-up the Skyrme EDF have been 
discussed in great detail in the 
literature~\cite{RMP,VB72,Per04a,papI,Sadoudi13,Ryssens14} 
and will not be recapitulated here.
Neither will we repeat the discussion concerning the different 
possible definitions for the surface energy. For a detailed discussion, 
we refer to the original articles of Myers and Swiatecki~\cite{MS80,MSW85}
and the more recent ones by Pearson \etal~\cite{FCP80,FP86}, 
Brack \etal~\cite{Brack85}, Kolehmainen \etal~\cite{koleh85}, 
Centelles \etal~\cite{Centelles98} or Douchin \etal~\cite{douchin00}.

%
%
\subsection{Hartree-Fock calculations}
\label{sect:hf}
The first one of the methods we use to calculate semi-infinite nuclear
matter is the self-consistent mean-field approximation, usually called 
Hartree-Fock (HF), with a treatment along the lines of 
Refs.~\cite{cote78,FCP80}. In this case, one considers 
the quantity~\cite{cote78}
\begin{equation}
\label{eq:quantity}
E_L
= \int_{-L}^{+L}\! \mathrm dz \; \mathcal{E}(z) \, ,
\end{equation}
which represents the energy per unit of surface for a piece of semi-infinite
nuclear matter described by a density which is constant in the $x$ and
$y$ directions and extends from $-L$ to $+L$ in the $z$ direction with
the conditions $\rho_0(-L)=\rho_\mathrm{sat}$ and $\rho_0(L)=0$ for 
$L\rightarrow +\infty$. After solving the mean-field equations,
the surface energy coefficient, denoted as $\asurf^\mathrm{HF}$, 
can be extracted using
\begin{equation}
\label{eq:esurf}
\asurf^\mathrm{HF} 
= \lim_{L\rightarrow\infty}\, 4\pi r_0^2
\int_{-L}^{+L} \! \mathrm dz \,
\Big[ \mathcal{E}(z) - a_v \, \rho_0(z) \Big] \, ,
\end{equation}
with the parameter $r_0$ being defined through the condition
$\frac{4}{3} \,\pi r_0^3 \rho_\mathrm{sat} = 1$.

As a particular feature of such quantal calculation, one observes so-called
``Friedel oscillations'' \cite{Friedel58,Ayachi87a}
of the density $\rho_0(z)$ in the vicinity of the surface inside the matter.
Since these oscillations are only very slowly damped, a reliable 
calculation of $\asurf^\mathrm{HF}$ requires quite large an
interval in $z$ direction.

Our HF values for $\asurf$ often differ slightly from those given by other 
groups in the past \cite{FCP81,Bar82a,farine97,Samyn05,samynprivate,Dan09a}, 
see the supplementary material \cite{supplement}. On the one hand, this 
underlines the numerical difficulties of determining a precise value for 
$\asurf$. On the other hand, as said before, the contribution from 
the $\vec{J}^2$ tensor 
terms, which are present for a subset of the parameterizations, has been 
omitted in most of the earlier published work. 

%
%
\subsection{Semi-classical calculations}
\label{sect:ETF}

As a second method to determine $\asurf$ we use the semi-classical 
Extended Thomas-Fermi (ETF) approach up to order
$\hbar^4$~\cite{Brack85}. Values obtained with this method will be
denoted as $\asurf^\mathrm{ETF}$ in what follows.

In this semi-classical framework, the local densities $\rho_q(z)$ are
modeled by a three-parameter modified Fermi function.
The kinetic and spin-current densities entering the total energy 
density $\mathcal{E}(z)$ are obtained from an expansion of the so-called
single-particle Bloch density matrix in powers of $\hbar$ around its 
Thomas-Fermi value as originally proposed by Wigner and Kirkwood. 
In short, the Bloch density matrix is the coordinate-space 
representation of the statistical operator $e^{\beta \hat{h}}$ constructed 
from the HF single-particle Hamiltonian $\hat{h}$ expressed in its eigenbasis 
\cite{Brack2001}, and related to the usual coordinate space representation 
of the density matrix $\rho(\vec{r},\vec{r}')$ through a Laplace 
transform \cite{Brack85,Brack2001}. Ultimately, this leads to 
expressions for $\tau_0$ and $\bJ_0$ as functions of the local density 
$\rho_0$ and its derivatives, where it is customary to separate the 
contributions of different (even) power in $\hbar$ in the expansion
\begin{eqnarray}
\tau_0 & = & \tau^{[0]}_0 + \tau^{[2]}_0 + \tau^{[4]}_0 \, ,
             \\
\bJ_0  & = & \bJ^{[2]}_0  + \bJ^{[4]}_0                 \, .
\end{eqnarray}
The complete expressions for $\tau^{[0]}_0$ (which is simply the 
kinetic energy of a non-interacting Fermi gas), $\tau^{[2]}_0$, 
$\tau^{[4]}_0$, $\bJ^{[2]}_0$ and $\bJ^{[4]}_0$ have been given 
by Brack \etal~\cite{Brack85} and more recently by Bartel 
\etal~\cite{Bartel02}.
Note that, in the case where tensor terms are included in the
Skyrme EDF, the standard expressions given in the early articles
have to be modified taking into account the results of Bartel 
\etal~\cite{Bartel08} for the contribution of the $\bJ^2$ terms.
Minimizing the surface energy as calculated with an expression 
equal to the one given by Eq.~\eqref{eq:esurf},
one obtains the parameters of the assumed Fermi-type density profiles
from which the surface energy coefficients can then be calculated.

%
%
\subsection{Modified Thomas-Fermi approximation}
\label{sect:MTF}
The ETF approximation provides expressions for the kinetic and 
spin-current densities in terms of the nucleon density and its derivatives.
The \textit{Modified Thomas-Fermi} (MTF) approximation developed 
by Krivine and Treiner~\cite{Krivine79} consists in using an ETF expansion
limited to order $\hbar^2$ where the coefficients are modified to simulate
the order $\hbar^4$ as well as the $\hbar^2$ effective mass contributions.
The MTF form of the kinetic and spin-current densities can then be written as
\begin{eqnarray}
\label{eq:mtf}
\tau^{[2]}_0  
& = & \alpha  k_F^2 \rho_0
      + \beta  \frac{(\vnabla \rho_0)^2}{\rho_0} 
      + \gamma \, \Delta \rho_0 
      + \tau^{[2(\textrm{so})]}_{0} , \\
\label{eq:mtf2}
\tau^{[2(\textrm{so})]}_{0}  
& = & \frac{1}{2} \, \bigg( \frac{2m_0^*[\rho_0]}{\hbar^2} \, \bW_0 \bigg)^2
      \rho_0  \, , \\
\label{eq:mtf3}
{\bJ}_0^{[2]} 
& = & -\frac{2m_0^*[\rho_0]}{\hbar^2} \, \rho_0 \, \bW_0  ,
\end{eqnarray}
with the coefficients
$\displaystyle\alpha=\tfrac{3}{5}$,
$\displaystyle\beta   = \tfrac{1}{18}$ and
$\displaystyle\gamma  = \tfrac{1}{3}$,
chosen for a resonable reproduction of the total energy of finite 
nuclei~\cite{Krivine79}. In local density approximation, the Fermi 
momentum is given by $k_{F} = (\frac{3}{2} \pi^2  \rho_0)^{1/3}$
and the spin-orbit field $\bW_0$ is defined as usual as
$\bW_0 = \partial \mathcal{E}/\partial \bJ_0$.

The coefficients $\alpha$ and $\gamma$ of the MTF expressions thus
keep their values obtained from ETF, whereas the value of $\beta$ is changed
from $1/36$ (obtained within ETF) to $1/18$ (used in the MTF method).

The interest of the MTF method is that in the case of symmetric semi-infinite 
matter, the approximations provided by Eqs.~\eqref{eq:mtf}, \eqref{eq:mtf2} 
and~\eqref{eq:mtf3} lead to an analytically solvable Euler-Lagrange equation
for the density profile without the need for carrying out a variational 
calculation numerically. It is noteworthy that the entire density
profile is varied in the MTF method, whereas in the ETF approach only
the parameters of a predefined Fermi function are optimized to give the
lowest binding energy.

Following the derivation outlined in Ref.~\cite{Treiner86}, the 
semi-classical energy density of symmetric matter can then be simply 
written as the sum of two terms
\begin{eqnarray}
\mathcal{E}
& = &   h_v[\rho_0]
      + h_s [\rho_0] \, \frac{\left( \vnabla \rho_0 \right)^2}{\rho_0} \, ,
\end{eqnarray}
where
\begin{eqnarray}
h_v[\rho_0] 
& = & C_0^\rho \rho_0^2 + B_0^\rho \rho_0^3 + D_0^\rho \rho_0^4 
               + \alpha \, \frac{\hbar^2}{2m_0^*} \, k_F^2 \, \rho_0  \,,
        \label{eq:hv} \\
h_s[\rho_0] 
& = & \frac{\hbar^2}{2m}\, \beta + d \rho_0 + g \rho_0^2
      + V_{\textrm{so}}[\rho_0] \rho_0^2 \,, \label{eq:hs}
\end{eqnarray}
with
\begin{eqnarray}
d 
& = & \left( \beta - \gamma \right) C_0^\tau - C_0^{\Delta \rho}   \,, \\
g 
& = & \left( \beta -2\gamma \right) B_0^\tau + B_0^{\nabla \rho}   \, , \\
V_{\textrm{so}}[\rho] 
& = & - \frac{1}{2} \frac{\left( C_0^{\nabla J} \right)^2}{Q[\rho_0]} \, .
\end{eqnarray}
Compared to what is found elsewhere in the literature, the 
expressions~\eqref{eq:hv} and~\eqref{eq:hs} have been complemented in order to 
take into account the possible 3- and 4-body contributions to the Skyrme EDF 
\eqref{eq:esk_sym} as introduced in Refs.~\cite{Sadoudi_Kaz12,Sadoudi13}. 
In addition, in order to take into account the tensor terms in the Skyrme EDF, 
the vector spin-current density is obtained through the protocol of Bartel
\etal~\cite{Bartel08}, which leads to a redefinition of the 
$V_{\textrm{so}}[\rho]$ term
\begin{eqnarray}
\bJ^{[2]}
& = & \frac{C_0^{\nabla J}}{Q[\rho_0]} \, \vnabla \rho_0 \, , 
      \\
Q[\rho_0]
& = & \frac{\hbar^2}{2m_0^*[\rho_0]} \, + \, C_0^J \, \rho_0
      \, + \, B_0^J \, \rho_0^2 \, ,
\end{eqnarray}
that, compared to the original work by Treiner and Krivine~\cite{Treiner86},
contains additional terms.

Inserting the analytical solution for the density profile $\rho_0(z)$ 
into the expression for the surface energy, one obtains after some 
further analytical manipulations a compact expression for the surface 
energy coefficient (for details see~\cite{Treiner86}), 
denoted $\asurf^\mathrm{MTF}$ from hereon,
\begin{equation}
\label{eq:pocket}
\asurf^\mathrm{MTF} 
= 8 \pi r_0^2 \, \int_{0}^{\rho_0} \! \! \! \mathrm d\rho \,  \,
    \bigg\{ h_s[\rho] \, 
            \bigg[ \tfrac{E}{A}(\rho) - \tfrac{E}{A}(\rho_{\text{sat}}) \bigg]
  \bigg\}^\frac{1}{2}
\, .
\end{equation}
As the MTF method differs from ETF approach only by the values of the 
coefficients, this expression has exactly the same form as the Wilets 
formula~\cite{Berg56,Wilets56} derived much earlier. Hereafter, we
will call Eq.~\eqref{eq:pocket} the \textit{MTF pocket formula} in order 
to underline the simplicity of its use to calculate $\asurf$ compared to 
the HF and ETF methods described above.

In its original MTF form without the contribution from the tensor terms,
this expression has been occasionally used to analyze the surface 
properties of Skyrme parameterizations
\cite{Treiner86,Margueron02a,Agrawal05}.

%
%
\section{Surface energy from semiclassical calculations of large nuclei}
\label{sect:nuclei}
%
%

In Ref.~\cite{Reinhard06}, Reinhard \etal\ have proposed a protocol 
to extract the nuclear liquid-drop coefficients that correspond to an 
EDF parameterization from a leptodermous expansion based on mean-field 
calculations of very large fictitious spherical nuclei calculated 
without Coulomb interaction and pairing correlations. 
For the determination of the (isoscalar) surface energy coefficient, 
the calculations can be limited to symmetric nuclei with $N = Z = A/2$. 
In this case, the calculated binding energy of the finite nuclei $E(A)$ 
can then be developed into
\begin{equation}
\label{eq:LDM:1}
E(A) 
\approx   \avol  \, A 
        + \asurf \, A^{2/3}
        + \acurv \, A^{1/3}
        + a_{0}  \, A^{0} \, ,
\end{equation}
i.e.\ a volume, surface, and curvature term plus another one that
is proportional to $A^0 = 1$. The latter has not been considered in 
Ref.~\cite{Reinhard06}. Despite its unrealistic limit for $A \to 0$, 
such term emerges naturally in the leptodermous expansion of the 
nuclear binding energy as a second-order correction to the curvature 
energy \cite{Pom03a,Roy06a}, or when replacing the geometric surface energy 
of the standard liquid-drop model by a double-folding integral 
\cite{Dav76a,Kra79a} as it is done in the finite-range liquid-drop 
and droplet models (FRLDM)~\cite{Mol95}.

For consistency with semi-infinite matter calculations, the 
finite nuclei are calculated without center-of-mass correction, 
irrespective of the scheme used during the fit of the parameterization 
used.

In order to extract the surface energy coefficient $\asurf$ of
Eq.~\eqref{eq:LDM:1}, one first defines an effective surface energy 
coefficient $\asurf^{\text{eff}}(A)$ of a given nucleus of mass
$A$ by reshuffling the expansion~\eqref{eq:LDM:1} 
\begin{eqnarray}
\label{eq:LDM:2}
\asurf^{\text{eff}}(A)
& \equiv & \bigg( \frac{E(A)}{A} - \avol \bigg) \, A^{1/3}
      \nn \\
& = &   \asurf 
      + \acurv \, A^{-1/3} 
      + a_0    \, A^{-2/3}
\, .
\end{eqnarray}
The volume energy coefficient $\avol$ is provided by the energy per 
particle $E/A$ of symmetric infinite nuclear matter at saturation 
density. By fitting a second-order polynomial in $A^{-1/3}$ to the 
calculated values for $\asurf^{\text{eff}}(A)$, one obtains the 
coefficients $\asurf$, $\acurv$ and $a_0$.

In order to disentangle the surface energy unambiguously from the 
higher-order terms in the liquid-drop formula \eqref{eq:LDM:1}, we 
calculate 20 nuclei with very large mass numbers in between 
$1 \, 200 \leq A \leq 200 \, 000$, similar to what was done
in Ref.~\cite{Reinhard06}. There, however, nuclei were calculated 
self-consistently, which required to remove shell effects by subtracting 
the shell correction as obtained from the self-consistent single-particle 
spectrum from the total binding 
energy. Here, we use the semi-classical ETF approach to calculate binding 
energies instead, such that there is no need to eliminate shell 
effects. The expansion in terms of powers of $A^{-1/3}$ (or inverse 
nuclear radius) is thus more stable and the LDM parameters thus more 
precisely extracted when extrapolating to (semi-)infinite nuclear matter 
in the limit $A^{-1/3} \to 0$. The such determined values for $\asurf$,
however, should be compared to the ETF results obtained for 
semi-infinite-matter, and not the HF results as the ones extracted
in Ref.~\cite{Reinhard06}.

%
%
\section{Determination of $\asurf$}
\label{sect:results}
%
%

In this section we present a systematic comparison of results for the
surface energy coefficient obtained with the four aforementioned methods. 
One of our main goals is to check whether the computationally friendly
MTF pocket formula provides 
a reliable estimate for the surface energy coefficient as extracted from 
the theoretically more advanced, but numerically more involved, HF or ETF 
methods. For that purpose, we use a large set of Skyrme parameterizations.

%
%
\subsection{Parameterizations considered}

There is a large number of Skyrme parameterizations that can be found in
the literature. Only few of them, however, are frequently used in 
production runs. The simplicity and popularity of the Skyrme EDF has 
led to quite large a number of ``experimental'' fits of parameterizations 
that were carried out for one or the other very specific study. 
In particular, there are many ``families'' of fits which explore the 
influence of variations of details of the parameterizations on their 
predictive power. We will profit here from this large number of 
parameterizations as it allows us to cover large intervals of values 
for infinite nuclear matter properties, which in turn can reveal
possible correlations of differences between the four methods to
determine $\asurf$ and other global features of the parameterizations.

Here we give a brief overview over the main features and particularities 
of the parameterizations considered here (which are all of the 
standard density-dependent bilinear form unless specified otherwise),
\begin{itemize}
\item 
SIII~\cite{Bei75a}: the coupling constants $C_t^\rho$ of this early, 
but still sometimes used, parameterization are linear functions of 
the density $\rho_0$; 

\item 
Ska~\cite{Kohler71}, SGI, SGII~\cite{Giai81}, SkM~\cite{skm}, 
SkM*~\cite{Bar82a}: these
are examples of early standard parameterizations with a density dependence 
with $\alpha < 1$, taking values of either $1/3$ or $1/6$ as almost all 
parameterizations listed below, unless otherwise specified;

\item 
The SLy family of fits: more recent and widely-used examples of standard 
parameterizations are the Saclay-Lyon fits SLy4-SLy7~\cite{Cha98a} that 
differ in options for center-of-mass correction and tensor terms, 
SLy5*~\cite{Pastore_Kaz12}, a recent refit that suppresses the finite-size 
spin-instability of the original SLy5 parametrization, and 
SLy4d~\cite{Kim97}, a refit of SLy4 without any center-of-mass correction 
built for the purpose of TDHF calculations;

\item 
The TIJ family of fits~\cite{papI}: for these parameterizations, also fitted 
within the Saclay-Lyon protocol, the coupling constants of the tensor terms 
were systematically varied over large intervals;

\item
SLy5+T \cite{Col07a}, SLy4T, SLy4T$_{\text{min}}$ \cite{Zal08a}, 
SLy4T$_{\text{self}}$, TZA \cite{papII}: these are further fits based 
on the Saclay-Lyon family of fits with either added or modified tensor terms;

\item 
f$_{0}$ and f$_{\pm}$ \cite{Les06a}: these constitute another 
series of variants of Saclay-Lyon parameterizations that explore different 
values for the splitting of the effective mass of protons and neutrons 
in asymmetric matter. Unlike the other standard parameterizations considered 
here, their coupling constants $C_t^\rho$ have \textit{two} density 
dependences with powers $\alpha_1 = 1/3$ and $\alpha_2 = 2/3$, 
respectively. This allows for the decoupling of nuclear matter properties 
that cannot be chosen independently for standard parameterizations with 
just one density dependence \cite{Cha97a};

\item 
The SLyIII.$x$ family of fits~\cite{Kouhei12}: these are yet another series 
of variants of Saclay-Lyon parameterizations that were built specifically 
for the purpose of regularized beyond-mean-field calculations. All have the 
same linear density dependence of the $C_t^\rho$ coupling constant as SIII. 
Their isoscalar effective mass $m^*_0/m=x$ has been constrained to the values 
0.7, 0.8, 0.9 and 1.0 in units of the nucleon mass;

\item 
SaMi~\cite{RocaMaza12}: this recent parameterization has been adjusted 
within a SLy-inspired protocol with the aim of improved spin-isospin 
properties;

\item 
SkI3, SkI4~\cite{Rei95a}, SkO, SkO'~\cite{Reinhard99}: these fits 
are examples of standard parameterizations with a generalized isospin 
dependence of the spin-orbit part of the EDF;

\item 
The SV family of fits~\cite{Klupfel09}: this series of parameterizations
was constructed for the purpose of studying the correlation of nuclear
matter properties with other observables. Starting from the reference 
parameterization called SV-bas, nine others were constructed by 
varying the incompressibility $K_{\infty}$, isoscalar effective mass 
$m^*_0/m$, symmetry energy $J$ and the sum-rule enhancement factor 
$\kappa_v$ one by one while keeping the others constant. A best fit,
called SV-min, is also considered;

\item 
The UNEDF family of fits: these parameterizations differ by the selection 
of data considered in the fit protocol. Compared to UNEDF0~\cite{Kortel10}, 
for example, UNEDF1~\cite{Kortel12} and UNEDF2~\cite{Kortel14} are also 
adjusted to reproduce the excitation energy of the fission isomer of 
\nuc{240}{Pu}.
We also consider the parameterization UNEDF1$^\textsc{SO}$~\cite{Shi14} 
that corresponds to a readjustment of the spin-orbit coupling constants 
of UNEDF1, thereby improving the spectroscopy of very heavy nuclei at
the expense of a much lower value for $\asurf$;

\item 
The BSk family of fits:
we also consider several representative parameterizations from the 
series of large-scale Bruxelles-Montreal Skyrme-HFB mass fits 
that stay within the standard form of the Skyrme EDF with only the
coupling constants $C^\rho_t$ being density dependent 
\cite{Samyn03,Gor03b,Samyn04,Samyn05,Gor05a,Goriely06,Goriely07,Goriely08}, 
among which BSk14~\cite{Goriely07} has also been fitted to fission barriers;

\item 
LNS~\cite{Cao06}, NRAPR~\cite{Steiner05}, NRAPRii~\cite{Ste13a}: in one way
or the other, these parameterizations of the standard Skyrme EDF were adjusted 
to reproduce nuclear matter properties as predicted by \textit{ab-initio} 
methods. While LNS was adjusted to reproduce a large variety of nuclear 
matter results from a Brueckner-HF calculation, NRAPR has been fitted to 
the density dependence of the energy per particle as obtained from 
\textit{ab-initio} calculations of nuclear and neutron matter. We also 
consider the parameterization NRAPRii with doubled strength $W_0$ of the 
spin-orbit interaction compared to NRAPR, as suggested by the authors 
of Ref.~\cite{Ste13a};

\item
KDE0v1 \cite{Agrawal05}: this standard parameterization has been adjusted 
to reproduce a large number of empirical nuclear matter data;

\item 
SQMC700~\cite{Guichon06}: this parameterization of the standard Skyrme 
EDF has been derived as the non-relativistic mean-field limit of a 
quark-meson-coupling model~\cite{Guichon06};

\item
S1Sd, S1Se \cite{Dob16a}: the parameters of these two standard 
density-dependent Skyrme parameterizations have been adjusted to 
reproduce total binding energies of doubly-magic nuclei as predicted 
by the Gogny force D1S. For S1Sd the tensor terms were included, 
whereas for S1Se they were neglected;

\item
S3Ly family of fits~\cite{Sadoudithesis}:
we also include a few representative examples from the series of 
fits of extended Skyrme EDFs that add central three-body terms 
with gradients to a standard density-dependent two-body Skyrme 
EDF, and which were carried out within a modified Saclay-Lyon 
fit protocol. These fits do systematically cover a wide 
range of values for the isoscalar effective mass $m^*_0/m$ and 
incompressibility $K_\infty$. For example, S3Ly71260 is a parameterization
with $m^*/m_0 = 0.71$ and $K_\infty = 260 \, \text{MeV}$;

\item
SLyMR0~\cite{Sadoudi_Kaz12}, SLyMR1~\cite{Robin_thesis}:
we also include two recent parameterizations built for the purpose of
spuriousity-free beyond-mean-field calculations: SLyMR0, 
which combines the non-density-dependent part of the standard 
2-body central and spin-orbit Skyrme interaction with gradient-less
3- and 4-body terms, and SLyMR1, where the 4-body 
terms are replaced by the 3-body terms with gradients as introduced 
in Ref.~\cite{Sadoudi13}.
\end{itemize}
The main interest of the four rarely-used parameterizations KDE0v1, LNS, 
NRAPRii and SQMC700 is that they were recently shown to be consistent 
with a large set of pseudo-data for infinite symmetric nuclear 
matter~\cite{Dut12a}. Their predictive power for finite nuclei, 
however, is rather limited \cite{Ste13a}.  Most, if not all, other 
parameterizations of the standard density-dependent Skyrme EDF listed 
above provide a much better description of finite nuclei than these, but 
in turn are incompatible with some of the presently accepted values for 
the empirical properties of nuclear matter~\cite{Dut12a}. It should be 
stressed, however, that none of the nuclear matter properties
is a real observable, as in one way or the other they all have to be 
extracted in a model-dependent way from data.

%
\begin{figure*}[t!]
\centerline{\includegraphics[angle=270,width=0.8\linewidth,bb=119 74 476 738,clip]{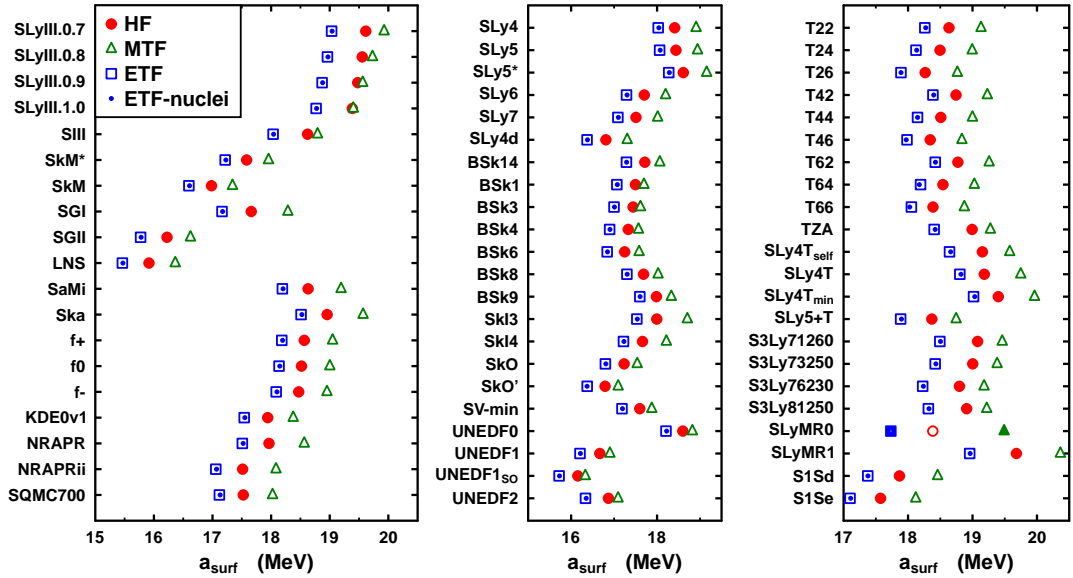}}
\caption{\label{fig:as_ref}
         (Color online)
         Surface energy coefficients obtained within HF, MTF and ETF 
         approaches for a large set of Skyrme parameterizations (see text).
         ETF values determined from semi-infinite matter calculations
         and extracted from calculations of finite nuclei as explained in
         Sect.~\ref{sect:nuclei}. 
         Note that all three panels share the same energy scale despite
         covering different intervals.
         Results for the SLyMR0 parameterization (inverted markers) have been 
	 artificially increased by three MeV in order to remain within the 
         range of the figure. 
}
\end{figure*}
%
%
\begin{figure*}[t!]
\centerline{\includegraphics[angle=270,width=0.8\linewidth,bb=119 74 476 738,clip]{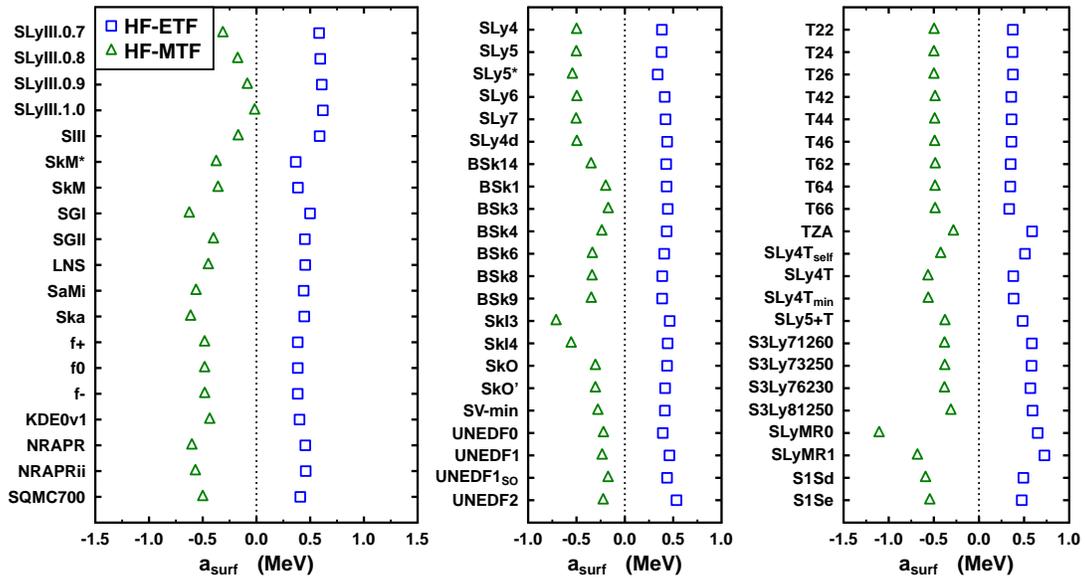}}
\caption{\label{fig:diff_ref}
         (Color online)
         Differences between the surface energy coefficients obtained
         with the HF, ETF and MTF models for the same parameterizations 
         as in Fig.~\ref{fig:as_ref}.
}
\end{figure*}
%

%
%
\subsection{Surface energy coefficients}
\label{sect:asurf:sinm}
Figure~\ref{fig:as_ref} shows the surface energy coefficients $\asurf$ 
obtained from the HF, ETF and MTF methods for the list of parameterizations 
given above. Two sets of ETF values, one obtained from the calculation 
of semi-infinite nuclear matter (open blue squares), the other extracted 
from calculations of large finite nuclei (blue dots), are shown.

The first observation that can be made is that the values of the surface 
energy coefficient spread over a relatively large interval, from about 
15.5\,MeV to about 19.5\,MeV for the results given by the HF calculations. 
As many parameterizations were obtained within dissimilar protocols, it 
is difficult to correlate the value for $\asurf$ with other properties 
of the respective parameterizations. A few correlations that can be 
unambiguously identified are that $\asurf$ depends on the presence and 
size of tensor terms (compare SLy4, SLy5, SLy5T, and the TIJ),
the size of the spin-orbit term (compare NRAPR and NRAPRii),
the scheme for center-of-mass correction used (as already pointed out 
in \cite{Ben00a}; compare SLy4, SLy6 and SLy4d).
One can expect that there are further correlations between a
parameterization's value for $\asurf$ and its other properties.
The analysis of their origin and nature, i.e.\ if they are a physical 
necessity or rather an accidental consequence of either a specific fit 
protocol or an over-constrained form of the EDF, however, is beyond 
the scope of the present study.

A second observation is that the differences between the values for 
$\asurf^\mathrm{HF}$ and $\asurf^\mathrm{ETF}$ on the one hand, and
the differences between $\asurf^\mathrm{HF}$ and $\asurf^\mathrm{MTF}$
on the other hand, are fairly constant and almost independent on the 
nature of the EDF considered (density-dependent bilinear, trilinear, 
or even quartic) and the properties it provides for infinite nuclear matter.

A third observation is that for all parameterizations the discrepancy
between the two different ETF values is so small that it can be hardly 
resolved on the Figure, which confirms that the leptodermous protocol
of Ref.~\cite{Reinhard06} offers a reliable alternative to calculations
of semi-infinite nuclear matter, with the remaining differences being
on the order of 10\,keV.

The size and parametrization-dependence of the little remaining scatter 
between the three methods to determine $\asurf$ from semi-infinite
nuclear matter calculations
can be better resolved on Fig.~\ref{fig:diff_ref}, where the 
differences $\Delta\asurf^{\mbox{\tiny HF-ETF}}
\equiv \asurf^\mathrm{HF}-\asurf^\mathrm{ETF}$ and 
$\Delta\asurf^{\mbox{\tiny HF-MTF}} \equiv 
\asurf^\mathrm{HF}-\asurf^\mathrm{MTF}$ are directly plotted
for the same sample of parameterizations as in Fig.~\ref{fig:as_ref}.
For most parameterizations, $\Delta\asurf^{\mbox{\tiny HF-ETF}}$ is close 
to $+0.5$\,MeV and $\Delta\asurf^{\mbox{\tiny HF-MTF}}$ is close to 
$-0.5$\,MeV. Nonetheless, in some cases the differences can deviate from 
this global trend; especially for $\Delta\asurf^{\mbox{\tiny HF-MTF}}$ one 
can find values in the range between $-1$\,MeV and about zero. In any 
event, Fig.~\ref{fig:diff_ref} indicates that, for the purpose of calculating 
$\asurf$, MTF is almost as good an approximation to HF as ETF. However, the 
two semi-classical methods differ among each other on a much larger scale 
as one systematically overestimates the HF value, whereas the other 
systematically underestimates it.

%
\begin{figure}[t!]
\begin{center}
\includegraphics[width=0.99\linewidth,bb=50 207 490 595,clip]{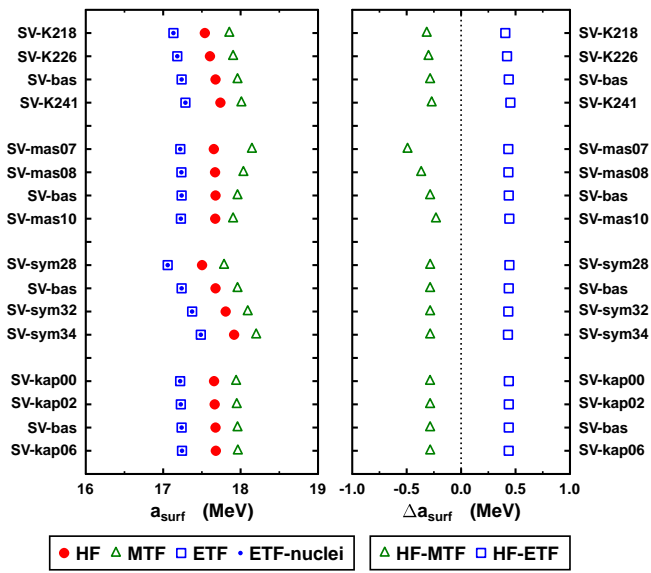}
\caption{(Color online)
Left: Surface energy coefficients obtained within the HF, MTF and ETF approach 
for the family of SV parameterizations of Ref.~\cite{Klupfel09}. Each of 
the four series systematically varies one bulk property (from top to bottom: 
$K_{\infty}$, $m^*_0/m$, $J$, $\kappa_v$) around the value of SV-bas while 
keeping the others constant. 
Right: Same as Fig.~\ref{fig:diff_ref}, but for the series of SV fits.
}
\label{fig:as_klupfel}
\label{fig:delta_as_klupfel}
\end{center}
\end{figure}
%

A closer examination of the parameterizations for which the scatter is 
largest indicates that $\Delta\asurf^{\mbox{\tiny HF-MTF}}$ may strongly 
depend on the isoscalar effective mass, see for example the SLyIII$.x$ and 
BSk series for which $m^\ast_0/m$ ranges between 0.7 to 1.05).

In some cases one also finds small differences between parameterizations 
with similar effective mass that are correlated to the strength of the 
tensor terms. This is exemplified by the comparison of SLy4, SLy5, 
SLy5+T and the T$IJ$ series that were adjusted within almost the same 
protocol, have practically the same effective mass, but differ in
the size and sign of the tensor coupling constants $C^J_t$. For these 
cases the difference between ETF and MTF seems not to be much affected, 
but that the differences between HF and ETF and also between HF and MTF 
are not the same. In general, Fig.~\ref{fig:diff_ref} suggests 
that without tensor terms, ETF and MTF have the same offset from HF, 
but in opposite directions. In the presence of tensor terms the ETF values
for $\asurf$ come closer to the ones from HF, whereas MTF moves further away.
This finding points to limitations of modeling the spin-current density
in semi-classical methods.

Comparing SLy5 and SLyIII.0.7, which have similar effective mass and 
tensor terms, indicates that $\Delta\asurf^{\mbox{\tiny HF-MTF}}$ might
in addition also depend on the power of the density dependence of
$C^\rho_0[\rho_0]$. For $\alpha = 1$ (SLyIII.0.7), MTF values for 
$\asurf$ are much closer to the ones from HF than for $\alpha = 1/6$ (SLy5). 
We will come back to this below.

One can expect that there might be further weak correlations between 
a parameterization's value for $\Delta\asurf^{\mbox{\tiny HF-MTF}}$ and 
its other properties, but these cannot easily be identified even within 
this large set of parameterizations. 

%
\begin{figure*}[t!]
\centerline{\includegraphics[angle=270,width=0.9\linewidth,bb=125 15 370 825,clip]{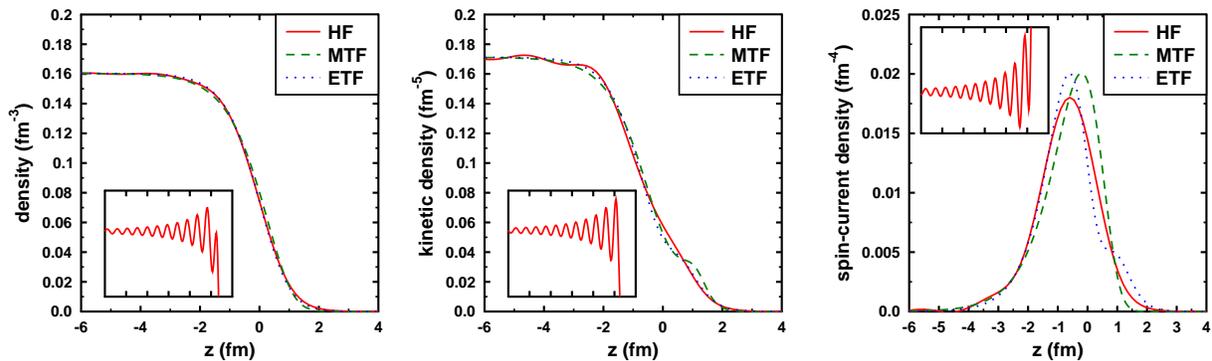}}
\caption{\label{fig:rho_tau_J_SLy5}
(Color online)       
Radial profile of the mass density $\rho(z)$, kinetic density $\tau(z)$, 
and the $z$-component of the spin-current density $J_z(z)$ as obtained
from HF, ETF and MTF calculations of semi-infinite nuclear matter with the
SLy5 parameterization. The inserts give an enlarged view of the Friedel
oscillations of the HF densities inside the matter in an interval of 
30\,fm below the surface.
}
\end{figure*}
%

In order to analyze further the correlation between 
$\Delta\asurf^{\mbox{\tiny HF-MTF}}$ and nuclear matter properties, 
parameterizations with systematically varied nuclear matter properties
are needed. Such fits are provided by the series of SV parameterizations 
by Kl{\"u}pfel \etal~\cite{Klupfel09}. 
During their adjustment, the incompressibility $K_{\infty}$, isoscalar 
effective mass $m_0^*/m$, symmetry energy coefficient $J$ and the 
Thomas-Reiche-Kuhn enhancement factor $\kappa_v$ have been separately 
varied while keeping the other properties constant. 

The left panel of Fig.~\ref{fig:as_klupfel} shows the corresponding values 
for $\asurf$ obtained with the four schemes to calculate the surface energy
coefficient introduced above. Two correlations for the size of $\asym$ itself, 
independent of the method of its determination, become immediately obvious: 
$\asurf$ increases with increasing incompressibility $K_\infty$ and
also with increasing symmetry energy coefficient $J$. Both, however, might
be particular to the specific fit protocol of this series of parameterizations.

More importantly, for the series with varied effective mass 
$m_0^*/m$, the different methods to determine $\asurf$ give a different 
trend, as already indicated by the analysis of the large set of 
parameterizations in Fig.~\ref{fig:diff_ref}. While the HF and ETF
values are fairly independent on $m_0^*/m$, the MTF values rapidly
increase with decreasing effective mass.
The method dependence of the $\asurf$ values becomes more obvious in
Fig.~\ref{fig:delta_as_klupfel}, where are plotted the differences
between the methods in the same way as in Fig.~\ref{fig:diff_ref}.

The right panel of Fig.~\ref{fig:delta_as_klupfel} also reveals that there 
is a slight dependence of both $\Delta\asurf^{\mbox{\tiny HF-ETF}}$ and 
$\Delta\asurf^{\mbox{\tiny HF-MTF}}$ on the incompressibility $K_\infty$, 
which to a large extent explains the difference of 
$\Delta\asurf^{\mbox{\tiny HF-MTF}}$ between SLy5 and 
SLyIII.0.7 found in Fig.~\ref{fig:diff_ref} and attributed to the power 
of the density-dependent term above. The values for $K_\infty$ of these 
two parameterizations are quite different, $K_\infty = 230$\,MeV for SLy5
and $K_\infty = 361.3$\, MeV for SLyIII.0.7, which is a consequence of the
correlation between $m^*_0/m$, $K_\infty$ and the power of the 
density-dependence of the $C^\rho_0$ coupling constant analyzed for 
example in Ref.~\cite{Cha97a}.
The $K_\infty$ value of SLyIII.0.7 is far outside the range covered by the 
SV parameterizations of Fig.~\ref{fig:delta_as_klupfel} (and therefore also
far from the empirical value, which is unavoidable for standard 
parameterizations with linear density dependence \cite{Cha97a}). 
Going from SV-K218 to SV-K241, one finds
that $\Delta\asurf^{\mbox{\tiny HF-MTF}}/\Delta K_\infty \approx 0.002$.
Assuming the same slope when going from SLy5 to SLyIII.0.7, then their
$\Delta\asurf^{\mbox{\tiny HF-MTF}}$ should differ by about 260\,keV, which
is indeed the case.

Altogether, we find a reasonably parameterization-independent behavior 
of the three models with a quantitative difference that is almost 
constant as long as $K_\infty$ and $m^*_0/m$ are constrained to a small
interval, which is usually the case within a given fit protocol. From this,
one can conclude that the Wilets pocket formula from Eq.~(\ref{eq:pocket})
can be safely used in a fit protocol to determine the coupling 
constants of future Skyrme EDFs.

%
%
\subsection{Density profiles in semi-infinite nuclear matter}
To examine the origin of the systematic differences found above
between the three models used to calculate semi-infinite nuclear matter,
Fig.~\ref{fig:rho_tau_J_SLy5} displays the radial profiles of the
mass density $\rho(z)$, kinetic density $\tau(z)$, and the 
$z$-component of the spin-current density $J_z(z)$ as obtained
from each method. 

For each model, the profiles are positioned such that the lower boundary 
$-L'$ of a piece of the surface with some sufficiently large, but otherwise 
arbitrarily chosen, particle number $A$ calculated as
\begin{equation}
\label{eq:sinm:surf}
A
= 4 \pi r_0^2 \int\limits_{-L'}^{\infty} \! {\rm d}z \; \rho(z) \, ,
\end{equation}
coincides for all of the three models.\footnote{None of our three codes
for HF, ETF, or MTF calculations of semi-infinite nuclear matter constrains 
the particle number in the integration interval. Still, within each model, 
the value of $-L'$ can be easily determined from the interpolation of $A(L)$
obtained from semi-infinite matter calculations with systematically
varied intervals $[-L,+L]$ in Eq.~\eqref{eq:quantity} etc. Indeed, 
for sufficiently large intervals, $A$ becomes a linear function of $L$. 
The value $A=40$ has been chosen to prepare Fig.~\ref{fig:rho_tau_J_SLy5}.}
The parameter $r_0$ is the same 
as in Eq.~\eqref{eq:esurf}. With this, the density profiles can be compared
exactly as those of finite nuclei with same particle
number. The origin $z = 0$, however, has been chosen to correspond 
to the position of the \textit{sharp} surface of a piece of saturated 
nuclear matter with constant density inside
that is placed in precisely the same manner as the three 
density profiles shown in Fig.~\ref{fig:rho_tau_J_SLy5}, i.e.\ 
$-L' = A/(4 \pi r_0^2 \rho_{\text{sat}}) 
     = A/[3^{1/3} (4 \pi \rho_{\text{sat}})^{2/3}]$.

As for finite nuclei, the density and kinetic density are bulk 
properties that approach a saturation value inside the matter and fall
off at the surface. By contrast, the spin-current density is peaked on the
surface and approaches zero inside and outside the matter. Indeed,
$J_z$ is exactly zero in infinite homogeneous nuclear matter when
calculated in mean-field approximation.

The amplitude of the Friedel oscillations exhibited by all three HF 
densities is comparatively small and barely visible when plotting the 
entire density profile. However, as indicated by the inserts, the 
oscillations with a wavelength of about 2.4\,fm are only slowly 
damped and reach far inside.

Each of the three approaches to calculate semi-infinite matter provides a 
slightly different profile of the mass density distribution $\rho(z)$
at the surface. They differ in the steepness, diffuseness, and also in 
the position of the inflection point. This is not very surprising, as in 
one way or the other the ETF and MTF methods are constructed to provide 
an approximation to the total energy of an HF calculation, and not to 
reproduce its densities. While in the ETF method only the parameters 
of a Fermi function parameterizing the profile of the local matter 
density $\rho(z)$ are variationally optimized, the entire density 
profile is implicitly varied in the MTF approach.

The ETF density $\rho(z)$ follows quite closely the mean of the oscillations 
of the HF density, such that the two are difficult to distinguish in 
Fig.~\ref{fig:rho_tau_J_SLy5}. By contrast, the profile of the MTF density 
is visibly different from the other two by being much more asymmetric around 
the surface. Inside the matter, the MTF density approaches the saturation 
value slower than the ETF density or the mean of the oscillating HF density, 
whereas outside the matter it falls much quicker to zero than the other 
two.  
The same behavior 
has already been found for finite nuclei in the seminal papers on the MTF 
method~\cite{Krivine79,Treiner86}. This asymmetry has some consequences 
for the corresponding kinetic and spin-current densities. On the one 
hand, it generates a visible bump in the tail of $\tau(z)$, and also leads
to a very asymmetric shape of $J_z(z)$.

A similar, but much less pronounced, bump is also found in the tail of the ETF
kinetic density $\tau(z)$. In this case, it is generated by the terms of 
order $\hbar^4$ in the semi-classical expansion, which are also at the
origin of the bump that the ETF spin-current density $J_z(z)$ exhibits 
at the same position. The appearance and size of the latter is 
parameterization-dependent, and can be correlated to the relative sign 
and size of spin-orbit and tensor terms.

While these differences in the density profiles easily explain why the
three methods deliver slightly different results for the surface energy 
coefficient, it is more difficult to correlate the dissimilarities of the 
density profiles with the systematic differences found for the $\asurf$ 
values. 
In any event, it has to be recalled that the ETF and MTF approaches are 
approximations to the HF calculation of the total energy of the nuclear 
system. The surface energy coefficient defined through Eq.~\eqref{eq:esurf},
however, is the difference between two energies that are typically two 
orders of magnitude larger, such that the differences found between the
various schemes to calculate them is beyond the third significant digit,
which should not be unexpected.

%
%
\section{Relation between $\asurf$ and
         deformation energy surfaces}
\label{sect:fis}
It is well understood that the characteristic features of the deformation
energy surfaces of heavy nuclei, i.e.\ the excitation energies of 
secondary minima and the height of barriers separating the minima 
and stabilizing the nucleus against fission, are strongly correlated 
with the surface properties of the effective interaction 
\cite{Bar82a,Nik11}. An emblematic example is the ``double-humped'' 
fission barrier of $^{240}$Pu \cite{Bjornholm80} that we will use 
here as illustrative example. As a rule of thumb, the larger the 
surface energy coefficient $\asurf$, the higher the 
excitation energies and barriers. 

In the liquid-drop model, the surface energy of a nucleus is simply
provided by the product of $\asurf$, the size
of the nucleus' surface and a universal factor. 
Assuming that the deformation of the fission isomer and the top
of the barriers turn out to be at the same deformation for all 
parameterizations (such that the nuclear surface is of comparable size),
one would then naively expect that their energy is linearly correlated 
with the surface energy coefficient. In addition, the larger the 
deformation, the steeper should be the slope.
	   
It has to be recalled, however, that the characteristic features of the
energy surfaces, in particular the ground-state deformation and the presence 
of secondary minima, are caused by shell effects \cite{funny,Bjornholm80}.
As a consequence, the variation of shell effects with deformation is
as important for the observable excitation energies as is the smooth 
variation of the of the liquid drop surface energy with deformation,
which on its own would just give all actinide nuclei a spherical 
ground state and one structureless broad fission barrier.
And indeed, the fact that the amplitude of the variation of shell effects 
with deformation that leads to the characteristic double-humped fission 
barrier of most actinides is correlated with the effective mass and 
spin-orbit strength has been pointed out already very early 
by Tondeur~\cite{Tondeur85}.

For a subset of the parameterizations employed above, 
we have carried out calculations of the complete static fission barrier 
of $^{240}$Pu, which is an often used benchmark for such studies
\cite{RMP,Bar82a,Berger91,Ben04a,Rutz95a,Ben00a,Bur04a,Younes09a,Schunck14a}.
In order to avoid a readjustment of the pairing strength and the ambiguities
related to it, we have limited the analysis to parameterizations of 
similar effective mass. Most were adjusted with a variant of the 
Saclay-Lyon protocol. 
In the figures we will only distinguish between series of fits,
which are the SLyx family, (SLy4-7, SLy5*, SLy4d), the fx family 
(f$_{0}$, f$_{\pm}$), the TIJ family (T22, T24, T26, T42, T44, T46, 
T62, T64), and the SLyxT family 
(SLy5+T, SLy4T, SLy4T$_{\text{min}}$, SLy4T$_{\text{self}}$, TZA). 
In addition, we have used the classic SkM and SkM* parameterizations.
As it turns out, this subset is sufficient for a conclusive analysis.

In all cases, we use ``surface pairing'' with strength 
$-1250 \, \text{MeV} \, \text{fm}^{-3}$ for protons and neutrons, 
and a soft cutoff at $\pm 5 \, \text{MeV}$ above and below the Fermi 
energy as defined in Ref.~\cite{Rigollet99}. 

Calculations are carried out with the most recent versions of the 
\textsc{Ev8}~\cite{Ryssens14} and \textsc{Ev4}~\cite{Ryssens_ev4} codes.
Both use the same 3D coordinate-space representation, where the single-particle
wave functions are discretized on an equidistant mesh in a box.
The two codes differ by the symmetries they impose on the nuclear shapes.
\textsc{Ev8} assumes three plane reflection symmetries, which reduces the
calculation to 1/8 of the full box, but is still sufficient to describe 
triaxial shapes. 
By contrast, \textsc{Ev4} assumes only two plane reflection 
symmetries, which then also permits to describe (not necessarily axial)
octupole deformed shapes. 
The \textsc{Ev8} box has $n_x \times n_y \times n_z = 20 \times 20 \times 30$ 
points of distance 0.8\,fm. The Poisson equation for the Coulomb field is 
solved in a $50 \times 50 \times 50$ 
box for improved precision of the Coulomb energy at large elongation. 
The \textsc{Ev4} box doubles the $z$ dimension.
With these choices, the deformation energy reaches a precision of 
about 100\,keV independent on the nuclear shape \cite{Rys15p}.

%
\begin{figure}[t!]
\includegraphics{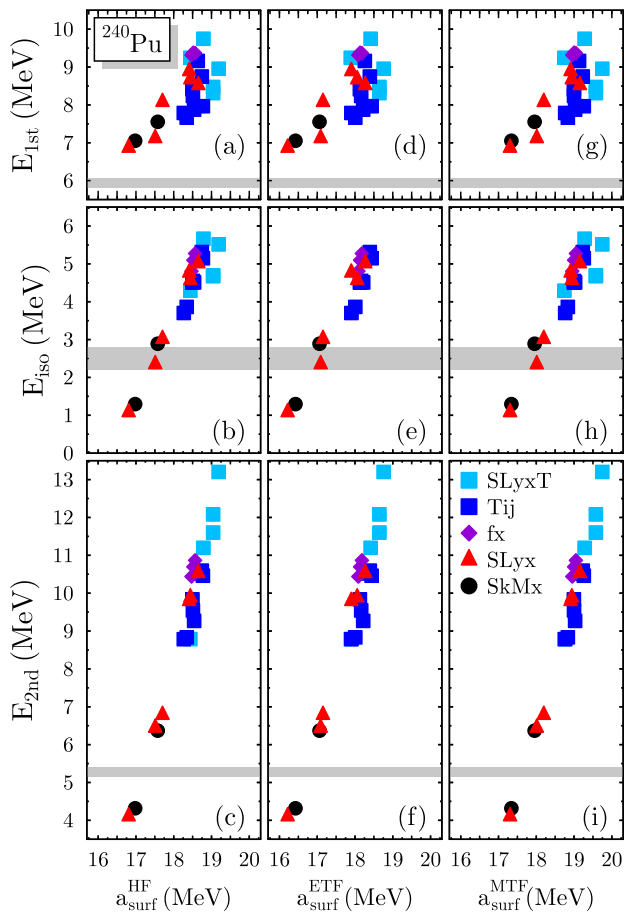}
\caption{\label{fig:as_isomer_all}
(Color online)
Correlation between the 
excitation energy of the fission isomer (middle panels),
the height of the inner (top) and outer (bottom panel) barrier
of $^{240}$Pu and the surface energy coefficient $\asurf$
calculated in HF (left column), ETF (middle column) and MTF (right column)
for a selection of Skyrme parameterizations (see text). 
The horizontal grey bars indicate the range of experimental data found 
in the literature (see text).
Note that all panels share the same energy scale.
}
\end{figure}
%

%
\begin{figure}[t!]
\begin{center}
\includegraphics{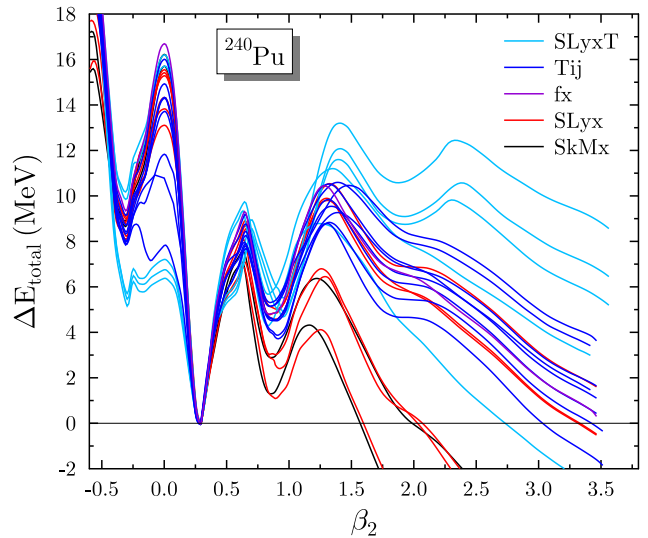}
\caption{\label{fig:pu240:all:def:1}
(Color online)
Deformation energy curve of \nuc{240}{Pu} as a function of the 
dimensionless mass quadrupole moment 
$\beta_2 = \frac{4 \pi}{3 R_0^2 A} \langle \hat{Q}_{2 0} \rangle$,
where $R_0 = 1.2 \, A^{1/3} \, \text{fm}$ and $A$ is the mass number,
for the same set of parameterizations as in 
Fig.~\ref{fig:as_isomer_all} and using the same color code for the 
families of parameterizations.
All deformation energies are normalized to the respective ground-state
energy. At large deformation, the curves end where the calculations
jump from a solution with large elongation to a solution with two separate
fragments.  
}
\end{center}
\end{figure}
%

Up to the fission isomer, the shapes along the static fission path 
are reflection symmetric, whereas beyond the fission isomer the
shapes quickly become mass-asymmetric with increasing quadrupole
deformation. The inner barrier is triaxial, everywhere else 
shapes along the path are axial.

Figure~\ref{fig:as_isomer_all} shows the excitation energy of the 
superdeformed fission isomer and the height of the inner and outer
barrier as a function of the surface energy coefficient calculated
in three different ways.

For the experimental energies, which are indicated by grey horizontal
bars in Fig.~\ref{fig:as_isomer_all}, slightly conflicting values have 
been reported in the literature. For the excitation energy of the isomer 
one finds 
$2.4 \pm 0.3 \; \text{MeV}$ \cite{Bjornholm80},
$\approx 2.8 \; \text{MeV}$ \cite{Singh02} 
and $ 2.25 \pm 0.20 \; \text{MeV}$ \cite{Hunyadi01}.
For the height of the barriers, the compilation of Ref.~\cite{Bjornholm80} 
lists $5.95 \; \text{MeV}$ for the inner and $\approx 5.4 \; \text{MeV}$ 
for the outer one.
According to the compilation quoted by Mamdouh \etal~\cite{Mamdouh98}, 
the first barrier has $5.8 \; \text{MeV}$ and the
second barrier $5.3 \, \text{MeV}$. 
A more recent paper from the same group \cite{Samyn05} quotes 
$<5.7 \; \text{MeV}$ for the inner and $4.5 \; \text{MeV}$ for 
the outer barrier. In their paper 
on the fit of UNEDF1, Kortelainen \textit{et al}.\ \cite{Kortel12} 
quote the values $6.05 \; \text{MeV}$ for the inner barrier and 
$5.15 \; \text{MeV}$ for the outer one, 
as recommended by the RIPL-3 database \cite{Capote09a,RIPL3}. 
The scatter of the experimental data, however, is much smaller than
the scatter in the theoretical results of Fig.~\ref{fig:as_isomer_all}. 
In any event, most parameterizations largely overestimate 
the experimental values, which is the principal motivation for 
the current efforts to improve the EDFs in that respect.

Note that none of the parameterizations does simultaneously describe
all of the three properties. The three parameterizations that give
a reasonable description of $E_{\text{iso}}$ of about 2.7\,MeV
(and which are SkM*, SLy6 and SLy7), still overestimate the inner and 
outer barrier heights by more than 1\,MeV.

All deformation energies increase with the value of the 
surface energy coefficient $\asurf$ as expected. 
However, the correlations are not strictly linear. Instead, there is 
a large scatter around the global trends, in particular for the 
height of the inner barrier.
This is an immediate consequence of the shell effects not being 
the same for all parameterizations. Indeed, it has been demonstrated in 
Ref.~\cite{Nik11} that the contribution of the shell correction as 
deduced from self-consistent mean-field calculations of the excitation 
energy of the fission isomer of nuclei in the actinide region, including 
\nuc{240}{Pu}, can vary by more than one MeV when going from
one parameterization to another. The parameterization-dependence of
shell effects becomes obvious when one directly compares the entire 
barriers, as can be seen from Fig.~\ref{fig:pu240:all:def:1}.
Even the overall shape of the barrier is not the same for all
parameterizations. The deformation of the fission isomer varies,
as does the deformation of the configuration corresponding to the
top of the barriers. 
Even more intriguingly, for some parameterizations there appears 
a third minimum at large asymmetric shapes, or the topography 
around the spherical point is quite different.
For the purpose of the present paper, it is not important
to disentangle the origin of these variations, which are for example 
related to the strength of spin-orbit and tensor terms as will
be discussed in a forthcoming article. 
What is relevant is that there obviously is a large variation of
the deformation dependence of the shell effects among the 
parameterizations studied here. This, in turn, indicates that the 
adjustment of the excitation energy of the fission isomer or the 
fission barrier heights cannot be easily replaced by an adjustment 
of a universal empirical value of the surface energy coefficient. 

%
%
\begin{table*}[t!]
\label{tab:SLy5_propb} 
\caption{Nuclear matter properties of the SLy5s$X$ Skyrme EDFs adjusted for
this work. The properties of the original SLy5 \cite{Cha98a} and 
SLy5*~\cite{Pastore_Kaz12} parameterizations are also given for comparison.
The first block (columns 2-8) shows the standard bulk properties for
infinite symmetric matter, i.e.\ saturation density $\rho_\mathrm{sat}$
in fm$^{-3}$,
energy per particle $E/A$ in MeV, incompressibility $K_\infty$ in MeV, symmetry
energy coefficient $J$ and its slope $L$ in MeV, and isoscalar effective
mass $m^*_0/m$ and enhancement factor of the TRK sum rule $\kappa_v$. 
The second block (semi-bulk; columns 9-11) shows surface 
energy coefficients $\asurf$ in MeV as obtained from semi-infinite nuclear 
matter calculations within the MTF, HF and ETF schemes (see text).
The third block (column 12) shows $\asurf$ as deduced from ETF 
calculations of large finite nuclei.
}
\begin{center}
\begin{ruledtabular}
\begin{tabular}{lcccccccccccc}
\noalign{\smallskip}
  & \multicolumn{7}{c}{bulk properties} 
  & \multicolumn{3}{c}{semi-bulk}
  & \multicolumn{1}{c}{from finite nuclei}                             \\
\noalign{\smallskip}\cline{2-8}\cline{9-11}\cline{12-12}\noalign{\smallskip}
 Model & $\rho_0$ & $E/A$ & $K_\infty$ & $m^*/m$ & $J$ & $L$ & $\kappa_v$ 
       & $\asurf^{\text{(MTF)}}$ & $\asurf^{\text{(HF)}}$
       & $\asurf^{\text{(ETF)}}$ & $\asurf^{\text{(ETF)}}$ \\
\noalign{\smallskip}\hline\noalign{\smallskip}
 SLy5   & 0.1603 & -15.98 & 229.9 & 0.6969 & 32.03 & 48.3 & 0.2498 & 18.94 & 18.44 & 18.06 & 18.07 \\
 SLy5*  & 0.1605 & -16.02 & 229.9 & 0.7006 & 32.01 & 45.9 & 0.4181 & 19.15 & 18.61 & 18.27 & 18.28 \\
 SLy5s1 & 0.1598 & -15.77 & 222.1 & 0.7392 & 31.43 & 48.1 & 0.3047 & 18.00 & 17.55 & 17.15 & 17.16 \\
 SLy5s2 & 0.1603 & -15.82 & 223.2 & 0.7350 & 31.60 & 48.3 & 0.3063 & 18.20 & 17.74 & 17.34 & 17.35 \\
 SLy5s3 & 0.1607 & -15.86 & 224.3 & 0.7309 & 31.77 & 48.4 & 0.3082 & 18.40 & 17.93 & 17.53 & 17.55 \\
 SLy5s4 & 0.1612 & -15.91 & 225.4 & 0.7273 & 31.94 & 48.5 & 0.3105 & 18.60 & 18.12 & 17.73 & 17.74 \\
 SLy5s5 & 0.1618 & -15.96 & 226.4 & 0.7243 & 32.11 & 48.6 & 0.3131 & 18.80 & 18.31 & 17.92 & 17.93 \\
 SLy5s6 & 0.1623 & -16.01 & 227.3 & 0.7217 & 32.29 & 48.8 & 0.3160 & 19.00 & 18.50 & 18.11 & 18.13 \\
 SLy5s7 & 0.1629 & -16.05 & 228.3 & 0.7196 & 32.46 & 48.9 & 0.3191 & 19.20 & 18.70 & 18.31 & 18.32 \\
 SLy5s8 & 0.1634 & -16.10 & 229.1 & 0.7178 & 32.64 & 49.0 & 0.3225 & 19.40 & 18.89 & 18.50 & 18.52 \\
\noalign{\smallskip}
\end{tabular}
\end{ruledtabular}
\end{center}
\end{table*}       
%

This is complicated further by the apparent impossibility of determining
a precise value for $\asurf$ in a model-independent way. 
As indicated by the analyses of Refs.~\cite{Reinhard06,Doba14},
a reliable extraction of $\asurf$ and other liquid-drop parameters
from microscopically \textit{calculated} binding energies of finite
nuclei requires to go to systems with $A \approx 10^{5}$ nucleons. 
The same can be expected to hold for its reliable extraction 
from the binding energies of real nuclei, but the systems needed do not 
exist in nature. Second, the preferred value for $\asurf$ will 
also depend on choices made for quantum corrections to the binding energy
in a given fit.
For example, the size of the rotational correction increases rapidly
with deformation \cite{RMP,Ben04a,Goriely07},
such that parameterizations that are supposed to be used with it require
a larger value for $\asurf$ than parameterizations supposed to
be used without it. Also, for parameterizations that are supposed to be
used in beyond-mean-field models, it is the energy difference between 
collective states that should be compared with data~\cite{Ben04a}, not 
the difference of minima of the energy surface.

%
%
\section{A fit protocol including $\asurf$}
\label{sect:sly5sx}

In one way or the other, information about fission barriers has already been 
sometimes used to constrain parameterizations of effective interactions 
of Skyrme \cite{Bar82a,Goriely07,Kortel12,Berger91} type.

Using the fission barriers themselves for the parameter adjustment, however, 
is quite costly. Assuming that shell effects are principally fixed by the 
other ingredients of the fit protocol, our results suggest a way to fit 
the information contained in fission barriers in a more economical iterative 
process:
A parametrization is first adjusted to reproduce an initial guess for the
value of $\asurf$, or a series of values of $\asurf$
that cover a reasonable range. After convergence of the fit, one calculates 
then a couple of well-selected fission barriers with the preliminary 
parameterizations, checks the deviation from data, estimates how
$\asurf$ should be changed and runs a new fit (or series of fits)
with the improved estimate for $\asurf$. The process can
be repeated until the desired quality for fission barriers is reached.

Figure~\ref{fig:as_isomer_all} suggests that the correlation between
the surface energy energy coefficient and the characteristic features 
of the energy surface is basically the same within each method to 
calculate $\asurf$, in spite of the HF, ETF and MTF calculations 
providing different values for a given parameterization. 
This means that any of these methods can then be used during the fit 
as long as the value used for $\asurf$ are tuned accordingly 
to the formalism used.

Because of its computational simplicity, we will use the MTF value
for $\asurf$ to construct a series of fits with a surface 
energy coefficient systematically varied in the range between 18.0 
and $19.4 \; \text{MeV}$ in steps of $0.2 \; \text{MeV}$, while
everything else in the fit protocol is kept unchanged. 
As a starting point, we chose the protocol used to adjust 
SLy5* \cite{Pastore_Kaz12}, which differs from the fit protocol 
of the original SLy5 parameterization~\cite{Cha98a} mainly by an 
additional constraint that avoids the appearance of unphysical
finite-size instabilities in the spin channels \cite{Pastore_Kaz12}.
In order to have a series of parameterizations with similar bulk
properties, we have added an additional constraint on the value of 
the $L$ coefficient of nuclear matter, which otherwise is only scarcely 
constrained by finite nuclei, such that it remains in the interval of 
$(50 \pm 2) \, \text{MeV}$.

\begin{figure}[t!]
\begin{center}
\includegraphics[width=0.7\linewidth,bb=150 324 443 600,clip]{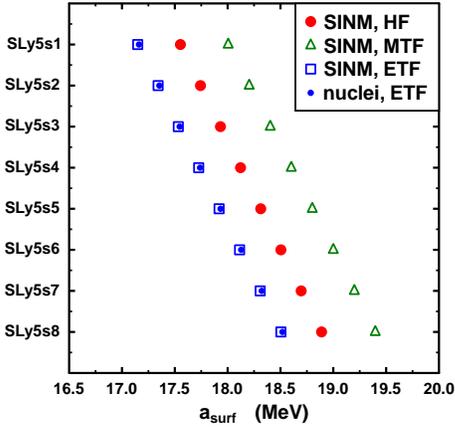}
\caption{\label{fig:as_KK}
Surface energy coefficients $\asurf$ of the
SLy5s$X$ parameterizations as obtained from HF, ETF and MTF 
calculations of semi-infinite nuclear matter as well as ETF
calculations of finite nuclei as described in Sect.~\ref{sect:nuclei}.
}
\end{center}
\end{figure}

The resulting parameterizations are called SLy5s$X$, $X=1$, \ldots, 8. 
Their nuclear matter properties are summarized in Table~\ref{tab:SLy5_propb}.
When comparing any two parameterizations, the difference between the 
values of the surface energy coefficients obtained with different methods 
is almost constant, see Fig.~\ref{fig:as_KK}.

%
%

%
\begin{figure}[b!]
\begin{center}
\includegraphics{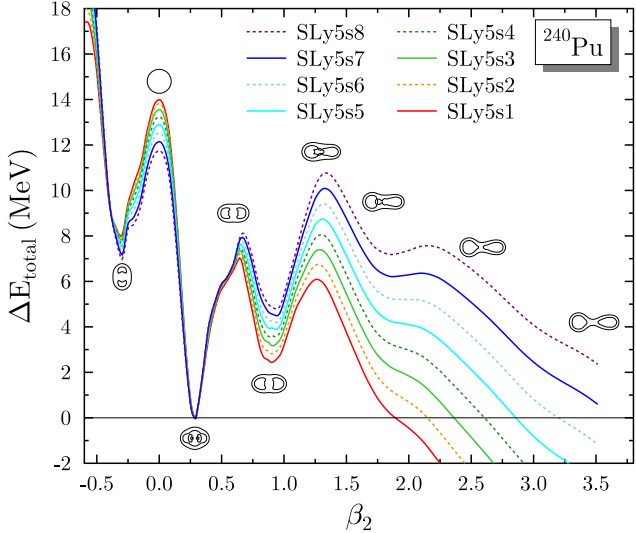}
\caption{\label{fig:pu_def}
(Color online)
Same as Fig.~\ref{fig:pu240:all:def:1}, but for the family of
SLy5s$X$ fits. The insets show contour plots of the mass density distribution 
in the $x$-$z$ plane at selected deformations.
}
\end{center}
\end{figure}
%

Figure~\ref{fig:pu_def} displays the deformation energy along the static
fission path of $^{240}$Pu obtained with these parameterizations. As they
are all fitted within the same protocol but for the value 
$\asurf$ is fixed at, the evolution of the curves with 
$\asurf$ is now much more regular.
However, there are still indications that also shell effects are slowly 
varying in response to the change of $\asurf$ in the fit protocol: 
at $\beta_2$ values around 1.7, a shallow third minimum develops 
with increasing value of $\asurf$. 

%
\begin{figure}[t!]
\begin{center}
\includegraphics{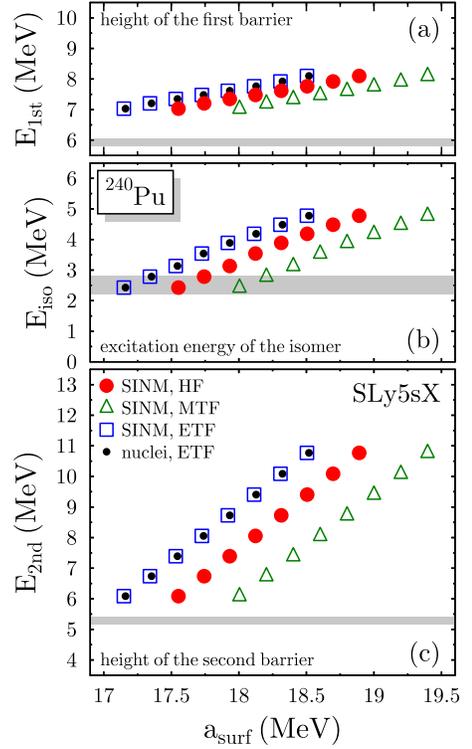}
\caption{\label{fig:as_isomer_SLy5xx}
(Color online)
Correlation between the excitation energy of the fission isomer (a),
the height of the inner (b) and outer (c) barrier
of $^{240}$Pu and the surface energy coefficient $\asurf$
calculated in HF, ETF, and MTF for the SLy5sx series of interactions.
The horizontal grey bars indicate the range of experimental data found 
in the literature (see text).
The energy scales and intervals are the same as in 
Fig.~\ref{fig:as_isomer_all}. 
}
\end{center}
\end{figure}
%

Still, for the excitation energies of the fission isomer and the inner and
outer saddles, the correlation with $\asurf$ is now almost perfectly
linear as demonstrated by Fig.~\ref{fig:as_isomer_SLy5xx}. Still, none of 
the parameterizations does simultaneously describe all of the three 
properties.

In the present study, we have focused on the isoscalar surface energy 
coefficient $\asurf$. All candidates for heavy nuclei whose deformation 
energy can be used to constrain parameter fits, however, have a neutron 
excess. For example, the nucleus $^{240}$Pu used as a benchmark above
has an asymmetry of 
$I \equiv (N-Z)/(N+Z) = 0.2$. The leptodermous expansion of the nuclear
binding energy suggests that there is a correction to the surface energy 
that depends explicitly on the nucleus' asymmetry $I$ and that in the 
liquid-drop model is parametrized through the surface symmetry energy 
coefficient $a_{\text{ssym}}$~\cite{Pom03a,Roy06a,Nik11,Reinhard06}. 

If a single nucleus is used to constrain the surface energy, there is 
the danger that $a_{\text{ssym}}$ accidentally takes a wrong value and
thereby introduces an erroneous isospin dependence of fission barriers.
However, as demonstrated in a detailed analysis of this quantity's 
role for the systematics of the excitation energy of superdeformed states 
of heavy nuclei~\cite{Nik11}, the value of the
$a_{\text{ssym}}$ of existing parameterizations of the Skyrme EDF is 
strongly correlated with the volume energy coefficient and therefore fixed
by masses along the valley of stability, such that it cannot vary freely 
over a wide interval. This gives us confidence that $a_{\text{ssym}}$ 
takes a reasonable and consistent value for all SLy5s$X$ fits. Indeed, for 
all of the SLy5s$X$ parameterizations we find values of $a_{\text{ssym}}$ 
that are close to $-49$\,MeV. Also,
calculations of the fission barriers of other heavy nuclei that will be 
reported elsewhere indicate that the overall trend of the predicted fission 
barriers does not change with asymmetry. For example, when calculated with
SLy5s1, the fission barrier height of the much less asymmetric nucleus 
\nuc{180}{Hg} with $I=0.11$ is described as well as is the one 
of \nuc{240}{Pu}.

%
%
\section{Discussion and Conclusions}
\label{sect:conclusion}

To summarize our main findings concerning the various methods to calculate
$\asurf$:
\begin{itemize}
\item[(i)]
HF, ETF and MTF calculations for semi-infinite nuclear matter
provide very consistent, but not identical, values for $\asurf$. 
The MTF method gives always slightly larger values than the HF approach, 
whereas the ETF method always gives slightly smaller values than HF. 
Differences between the values extracted from semi-classical MTF and ETF 
calculations on the one hand and the quantal HF approach on the other 
hand are typically on the order of 500\,keV out of about 18\,MeV,
but in exceptional cases might be as large as 1\,MeV. 

\item[(ii)]
The values of $\asurf$ as extracted from ETF calculations of semi-infinite 
nuclear matter and from the systematics of the ETF binding energies of very 
large artificial spherical nuclei are very close and differ rarely by more
than 10\,keV. To reach this level of agreement, an $A^0$ term has to be
included in the leptodermous expansion of the binding energy of finite 
nuclei, and the contribution of the center-of-mass correction omitted.

\item[(iii)]
For the purpose of calculating $\asurf$, the MTF and ETF approximations are 
fairly robust. The deviation from HF values is reasonably parameterization 
independent, although there are differences in detail that are correlated to 
the presence and strength of the tensor terms, the incompressibility 
$K_\infty$, and the isoscalar effective mass $m^*_0/m$. During a fit, 
the values for $K_\infty$, $m^*_0/m$ and the coupling constants 
of the tensor terms will rarely vary over a large interval, such that the 
analytical MTF value for $\asurf$ as obtained from from Eq.~(\ref{eq:pocket})
can be safely used in a fit protocol to determine the coupling 
constants of Skyrme EDFs. 
In any event, we do not aim at the reproduction of a universal 
empirical value for $\asurf$, which will be difficult to extract in a 
model-independent way from data anyway, but instead provide a simple and 
efficient control over the surface properties within a given framework.

\end{itemize}
Concerning the correlation between characteristic energies of the
fission barrier of \nuc{240}{Pu} and the values for $\asurf$ we find that 
\begin{itemize}
\item[(i)]
The simultaneous description of the fission barrier heights and the 
excitation energy of the fission isomer is not trivial as they are
also strongly sensitive to shell effects which are at the very origin 
of the complicated topography of the deformation energy surface.

\item[(ii)]
While the characteristic energies of fission barriers are clearly and 
unambiguously correlated to the surface energy coefficient in the 
expected manner, for existing Skyrme parameterizations there is a 
large scatter when plotting them as a function of the surface energy 
coefficient. This is a consequence of the shell effects being 
unsystematically different when parameterizations are constructed with 
different fit protocols. 

\end{itemize}
To eliminate the protocol-dependence of the analysis of $\asurf$ and
its correlation with fission barriers, and as a proof of principle for
a parameter fit that includes the MTF value for $\asurf$ in its protocol,
we have constructed a series of parameterizations of the standard Skyrme 
functional called SLy5s1, SLy5s2, \ldots, SLy5s8, with systematically 
varied surface energy coefficient. 
\begin{itemize}
\item[(i)]
The resulting parameterizations exhibit almost linear correlations
between the surface energy coefficient on the one hand and each of the 
characteristic energies of the energy surface of \nuc{240}{Pu} on the 
other hand, providing the proof-of-principle for replacing the adjustment
of fission barriers with a suitably chosen value of $\asurf$ as obtained
from semi-infinite nuclear matter calculations with any of the schemes
considered here.

\item[(ii)]
None of the SLy5s$X$ parameterizations, however, reproduces simultaneously 
the empirical data for the excitation energy of the fission isomer as 
well as the heights of the inner and outer barriers. This points to 
imperfections of the deformation-dependence of shell effects provided by
these parameterizations.

\item[(iii)]
The best reproduction of the energy surface of \nuc{240}{Pu} is provided
by SLy5s1, the parameterization with the lowest value of $\asurf$ 
considered in the series of fits.

\end{itemize}
Adjusting parameters to reproduce a value for $\asurf$ is computationally 
much easier than reproducing fission barriers of heavy nuclei, not only with
respect to CPU time, but, even more importantly, also in terms of stability 
and controllability of the calculations during the early stages of a fit when 
the parameters are still far from their physical values such that the 
topography of the energy surfaces might be very different from the physical 
one. 
In practice, however, the appropriate value of $\asurf$ that provides 
the best description of the targeted deformation energies will have to 
be determined iteratively, alternating between complete parameter fits
for some guess(es) for $\asurf$ on the one hand, and the calculation of 
deformation energy surfaces with the such determined parameterizations 
on the other hand. The latter then provide a refined guess for $\asurf$, 
if necessary. In the present paper, we limited ourselves to the first 
of such cycles.

As many of the widely-used Skyrme parameterizations provide an unsatisfactory
description of the systematics of fission barriers and excitation energies
of very deformed states \cite{RMP,Nik11,Bur04a,Erler12a}, it is highly 
desirable to better constrain the nuclear surface energy in future fits. 
Besides the obvious importance for the study and understanding of the 
fission process itself, fission barrier heights are also a determining 
factor for the stability of superheavy nuclei \cite{Bur04a,Erler12a} and 
the dynamics of the astrophysical r-process of nucleosynthesis 
\cite{Samyn05,Goriely07,Erler12a}.

In conclusion, the use of the MTF pocket formula for $\asurf$ provides 
a rapid and robust expression to control the surface properties of
nuclear energy density functionals during the adjustment of their 
parameter.
In this paper, we have focused on the (isoscalar) surface energy
coefficient. Within the MTF approach, the integral that appears in 
the calculation of the (isovector) surface \textit{symmetry} energy 
coefficient from semi-infinite matter cannot be solved analytically.
However, making further approximations whose consequences 
remain to be analyzed, one can arrive at an analytical expression for
this quantity as well \cite{Krivine1983}, which might offer a route
to its efficient fine-tuning.

The generalization of the MTF pocket formula to Skyrme-type functionals 
with derivative terms of order 4 or 6~\cite{Raimondi11,Dav13a,Bec15a} 
is feasible, but will require an arduous extension of the formalism to 
higher-order terms.

%
\section*{Acknowledgments}

Stimulating discussions with D.~Davesne about semi-classical methods 
in physics are gratefully acknowledged. We also thank W.~Ryssens for 
critical reading of an early version of the manuscript.
This work was supported by the Agence Nationale de la Recherche
under Grant No.\ ANR 2010 BLANC 0407 ``NESQ'',
by the CNRS/IN2P3 through PICS No.~6949,
and by the Academy of Finland and University of Jyv\"askyl\"a
within the FIDIPRO programme. 
Part of the computations were performed using HPC resources of the MCIA 
(M{\'e}socentre de Calcul Intensif Aquitain) of the Universit{\'e} de 
Bordeaux and of the Universit{\'e} de Pau et des Pays de l'Adour.

%
%


\bibliography{jodon_surface_energy.bib}


\end{document}